\documentclass[aip,apl,twocolumn,reprint,amsmath,amssymb]{revtex4-1}
\usepackage{graphicx}
\usepackage{textcomp}
\usepackage{upgreek}
\usepackage{braket}
\usepackage{color}

\begin{document}
\title{Tapering of fs Laser-written Waveguides} \normalsize

\author{Ren\'{e} Heilmann}
	\affiliation{Institut f\"ur Physik, Universit\"at Rostock, Albert-Einstein-Stra\ss e 23, 18059 Rostock, Germany}
	\affiliation{Institute of Applied Physics, Abbe Center of Photonics, Friedrich-Schiller-Universit\"{a}t, Max-Wien-Platz 1, 07743 Jena, Germany.}
\author{Chiara Greganti}
	\affiliation{Faculty of Physics, University of Vienna, Boltzmanngasse 5, A-1090 Vienna, Austria.}
\author{Markus Gr\"afe}
	\affiliation{Institut f\"ur Physik, Universit\"at Rostock, Albert-Einstein-Stra\ss e 23, 18059 Rostock, Germany}
	\affiliation{Institute of Applied Physics, Abbe Center of Photonics, Friedrich-Schiller-Universit\"{a}t, Max-Wien-Platz 1, 07743 Jena, Germany.}
\author{Stefan Nolte}
	\affiliation{Institute of Applied Physics, Abbe Center of Photonics, Friedrich-Schiller-Universit\"{a}t, Max-Wien-Platz 1, 07743 Jena, Germany.}
\author{Philip Walther}
	\affiliation{Faculty of Physics, University of Vienna, Boltzmanngasse 5, A-1090 Vienna, Austria.}
\author{Alexander Szameit}
\affiliation{Institut f\"ur Physik, Universit\"at Rostock, Albert-Einstein-Stra\ss e 23, 18059 Rostock, Germany}
	\affiliation{Institute of Applied Physics, Abbe Center of Photonics, Friedrich-Schiller-Universit\"{a}t, Max-Wien-Platz 1, 07743 Jena, Germany.}

\bigskip
\begin{abstract}
The vast development of integrated quantum photonic technology enables the implementation of compact and stable interferometric networks. In particular laser-written waveguide structures allow for complex 3D-circuits and polarization-encoded qubit manipulation. However, the main limitation for the scale-up of integrated quantum devices is the single-photon loss due to mode-profile mismatch when coupling to standard fibers or other optical platforms. Here we demonstrate tapered waveguide structures, realized by an adapted femtosecond laser writing technique. We show that coupling to standard single-mode fibers can be enhanced up to 77\% while keeping the fabrication effort negligible. This improvement provides an important step for processing multi-photon states on chip.
\end{abstract}

\maketitle

Photonic chip technology has become a widely used platform when it comes to solid-state simulation~\cite{Szameit2010}, sensor applications~\cite{Crespi2012}, or quantum computing~\cite{OBrien2009}. Especially the latter field shows a deep interest into integrated device, since photonic quantum architectures promise high complexity, miniaturization, and long term stability~\cite{Politi2008,Politi2009,Matthews2009,Crespi2011}. So far, most of the current quantum optics experiments are realized by utilizing different stages, such as bulk optics for state preparation, fiber networks for guiding, and integrated waveguide structures for quantum operations. Hence, one major part of experimental setups dealing with integrated waveguide is the in and out coupling of the vital radiation. Most challenging is the matching of mode shapes provided by different optical systems, for instance single mode fiber and optical chip. Free space coupling offers a suitable opportunity for altering the mode profile, but suffers from extensive alignment effort and much larger footprint size as well as weak phase stability compared to integrated solutions. Furthermore, complex interface structures that involve more than one channel demand arduous tailored designs and substantial experimental costs~\cite{Keil2011,Kondakci2016}.
\par
Femtosecond (fs) laser-written photonic waveguide structures~\cite{Meany2015} additionally offer three-dimensional (3D) waveguide designs. Photonic input and read out in such intricate geometries is conveniently realized by 
butt coupling of the optical waveguides: in case of fibers these are directly attached to the waveguides in the optical chip. In doing so, it is possible to fulfill the request for multi-channel connections as used in boson sampling experiments~\cite{Tillmann2013,Crespi2013-2} and in a variety of enormous readout tasks regarding W-state generation~\cite{Graefe2014}, integrated photonic quantum walks~\cite{Graefe2016}, quantum Bloch oscillations~\cite{Lebugle2015}, and integrated quantum Fourier transforms~\cite{Weimann2016,Crespi2016}. On the downside, the coupling efficiency was considered a huge obstacle since it prevents to minimize the overall losses that are especially crucial for multi-photon experiments.
\par
In essence, the mode field diameter (MFD) of fs laser-written waveguides in fused silica exceeds the standard fiber modes by a factor of two~\cite{Keil2016} resulting in a coupling efficiency falling below 50\%. Instead of using standard single mode fibers one could use large mode area fibers bringing the coupling efficiency up to 90\%. On the contrary, such fibers are rather expensive, require scrupulous dealing concerning bending radii, and are so far not suitable for fiber arrays with small pitches.
\par
Here, we present a new method to shrink the mode size of the waveguides on an optical chip using the established fabrication technique for laser inscription. By introducing a tapered region close to the facets the advantages of integrated waveguide structures remain unaffected. We take a closer look on the different impact of power modulation and repetitions of the writing process on the final mode size. Thereupon, a most effective writing scheme is proposed that does not need more than a few millimeters of additional space. Finally the performance is investigated by comparing the enhanced coupling efficiency to the occurring transmission losses.
\par
The waveguides were fabricated using fs laser pulses delivered from a Ti:Sa oscillator with a repetition rate of 100\;kHz. In a moderate pulse energy regime from 200 till 500 $\upmu$J the absorption in fused silica at the focal point leads to a permanent refractive index change with a contrast of up to $10^{-4}$~\cite{Davis1996,Meany2015}. By moving the glass sample with respect to the laser focus waveguides of any trajectory can be drawn. Low magnification objectives (20$\times$) are preferred to achieve a working distance of several millimeters allowing to write 3D structures~\cite{Hnatovsky2005,Heilmann2015}.
\par
In an earlier attempt to decrease MFDs in fs laser-written waveguides, beam shaping with a spatial light modulator was applied~\cite{Salter2012}. While this approach is very attractive and versatile, it goes hand in hand with extended experimental effort, low fabrication rates, and most importantly, it disables the 3D capability, since waveguides will only be identical in a fixed plane (or in a very narrow stripe of few tens of $\upmu$m). Another work demonstrated the adiabatic conversion of a waveguide supporting higher-order modes into one supporting only one single mode~\cite{Gross2014}. This was achieved by linearly changing the pulse energy during the inscription process. Additional annealing of the optical chip was necessary to obtain optimized performance. 
\par
\begin{figure}[t]
	\includegraphics[width=\linewidth]{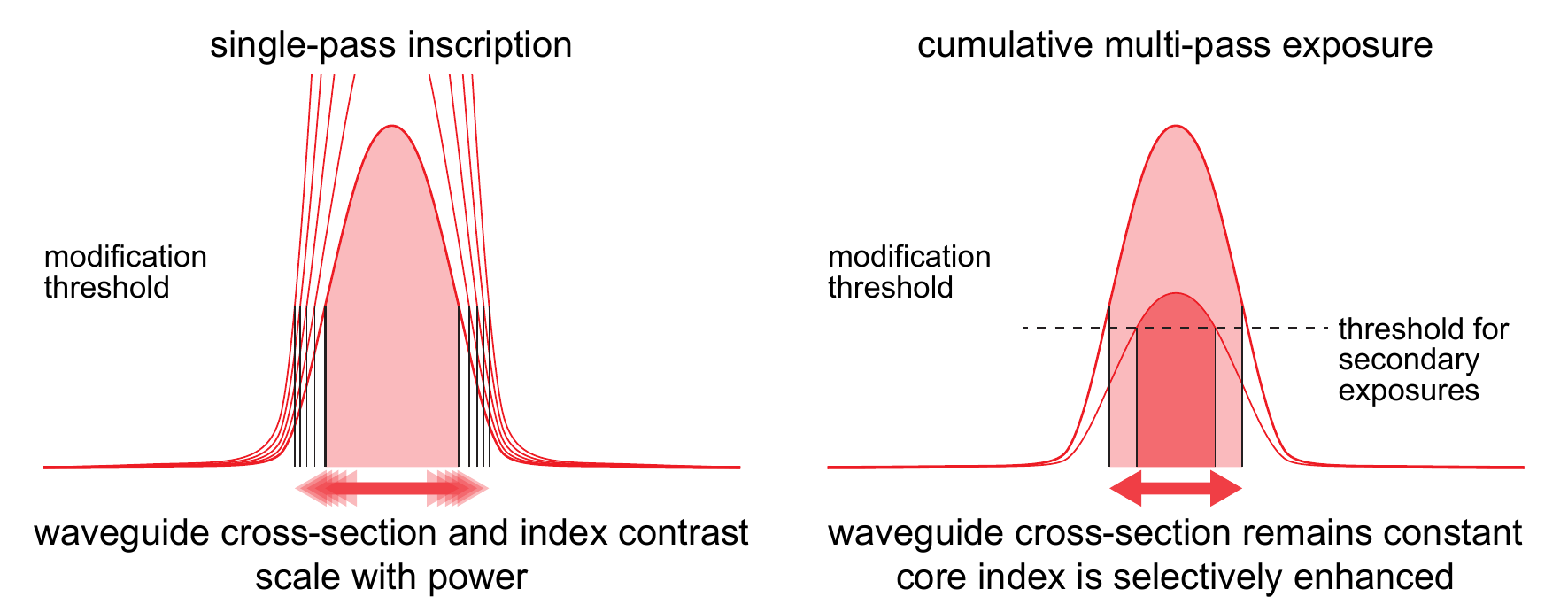}
	\caption{Left: Illustration of the dependency of the waveguide cross-section on the inscription power. Higher power levels increase the refractive index contrast, however, also extend the modification volume which suppresses efficient mode compression. Right: Multiple reruns after the initial inscription with lower power level will also increase the refractive index while keeping the size of the modified region constant. The modification threshold for the multiple reruns is slightly below the one from the first inscription process.}
	\label{fig:multipass}
\end{figure}
\begin{figure}[t]
	\includegraphics[width=\linewidth]{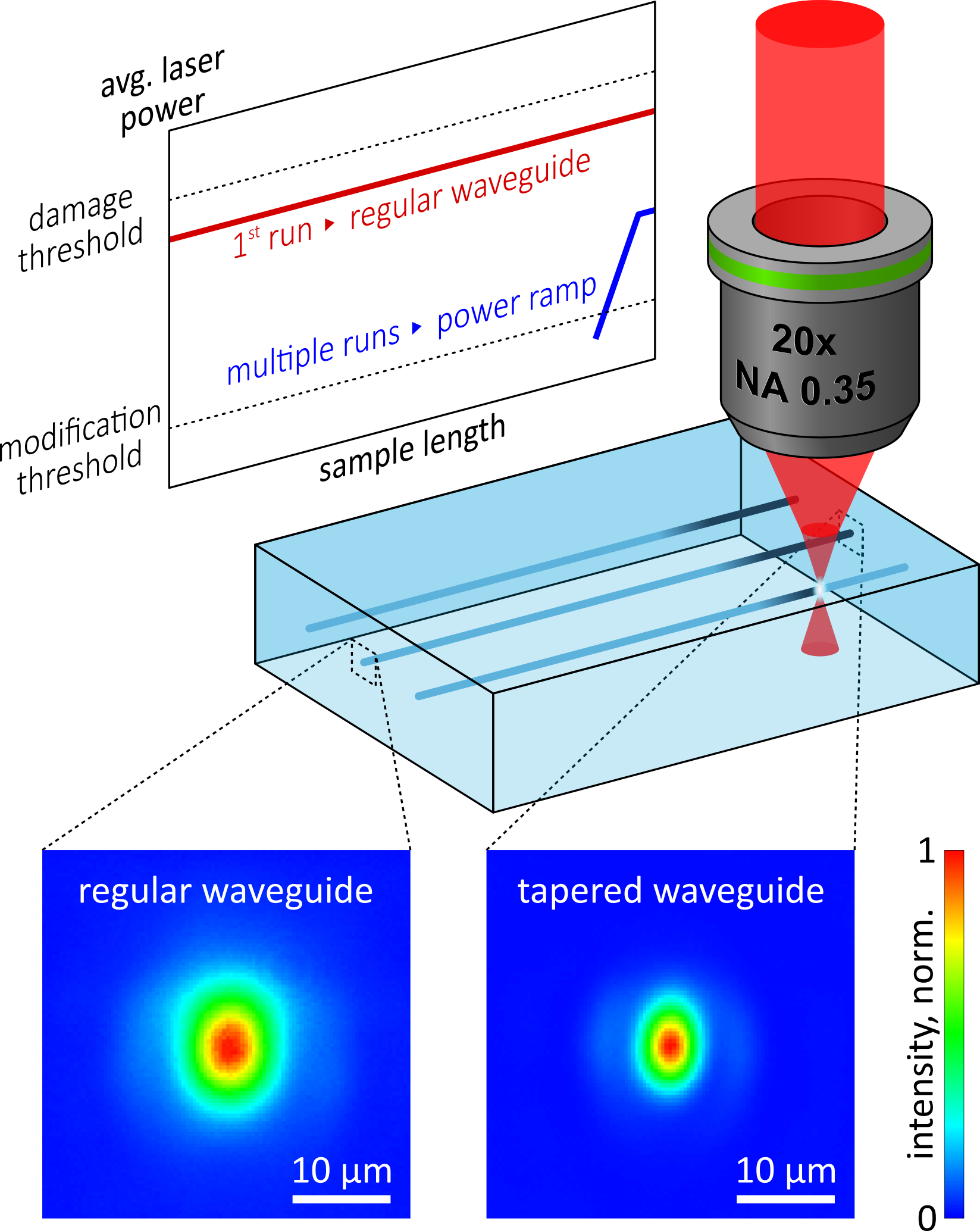}
	\caption{Top: Illustration of the fabrication process of tapered waveguides. Using the three dimensional fs laser-writing technique a regular waveguide is inscribed in the first place with a power of $P_0$. Secondly, close to the facet of the glass chip additional runs are performed with increasing laser power starting from right under the modification threshold up to a moderate power level $P_\mathrm{a}$. Bottom: The near field images of the guided modes for a regular and tapered waveguide prove the adiabatic change of the guided mode towards smaller mode sizes.}
	\label{fig:setup}
\end{figure}
The last strategy can by adapted but needs to be significantly modified for the purpose of lowering the MFD of single mode waveguides. Taking the ansatz of Ref.~23, a spatial power ramp for inscribing the last few millimeter of a waveguide will increase the refractive index but at the same time enlarge the modification region (see Fig.~\ref{fig:multipass}), which counters the desired mode size shrinking. Our intent is to increase the refractive index while maintaining the modification volume. This can be realized by multiple reruns over the last few millimeter of the waveguide with a spatial power ramp exhibiting a maximum power equal or below the initial inscription power (see Fig.~\ref{fig:multipass} \& \ref{fig:setup} top).
\par
In doing so, we keep the effort in the fabrication process minimal and maintain the 3D writing capability. 
In detail, the reruns started with an average laser power below the modification threshold 3\;mm from the end of the glass sample and followed a linear power ramp up to a moderate level as the focal point approaches the edge. Both, the number of reruns and the corresponding power levels have been tested.
We wish to emphasize that the input site was identical for all examined waveguides. Tapered regions were only implemented at the output sites where the impact on the MFD is clearly observable (see Fig.~\ref{fig:setup} bottom). The input coupling was based on a 10$\times$ microscope objective being adjusted by a three axis piezo translation stage to ensure the same input conditions from waveguide to waveguide. At the output side the mode fields where imaged by a 10$\times$ objective onto a CCD camera.
\par
\begin{figure}[t]
	\includegraphics[width=\linewidth]{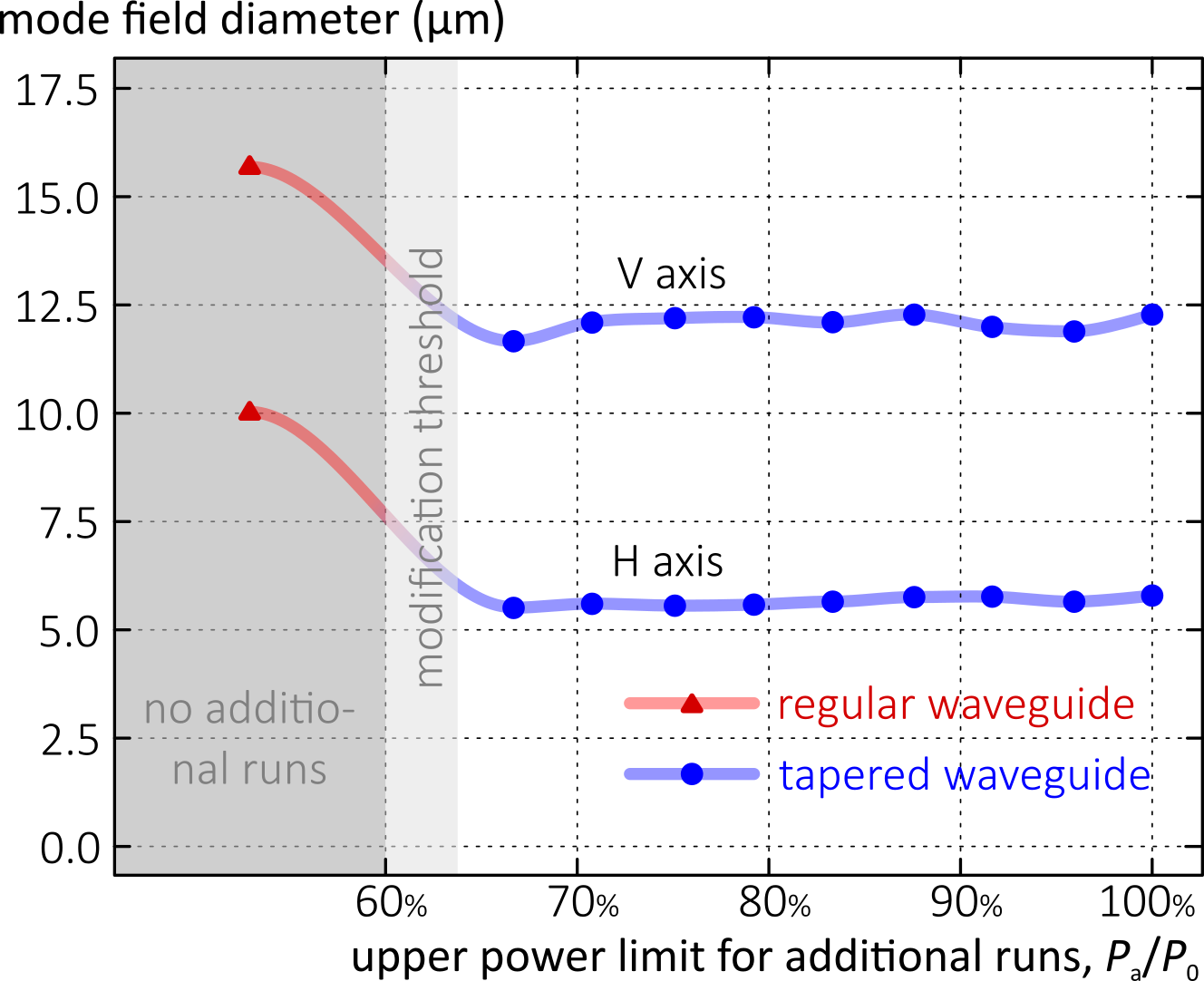}
	\caption{Modification of the mode field diameter ($1/\mathrm{e}^{2}$ width in intensity) for horizontal (H) and vertical (V) axis at 800\;nm with 16 additional runs along the tapered region. The upper power ramp limit $P_\mathrm{a}$  of the additional runs was tuned from $66.7\%$ to $100\%$ of $P_0$ for a regular waveguide inscription. Compared to a regular waveguide (red triangles), tapered waveguides (blue dots) show a significant reduced MFD. The width of the associated lines represents the error regime of $\pm1.3$\;nm.}
	\label{fig:power}
\end{figure}
In a first experiment the number of repetitions was set to 16. The upper laser power limit $P_\mathrm{a}$ during the additional runs was increased from slightly above the modification threshold up to the power level $P_0$ used for the regular waveguide, which was safely below the damage or multi mode threshold for the inscription process.
In Fig.~\ref{fig:power} the MFDs for different power limits are separately shown for the horizontal and vertical axis.
On both axes the MFD is reduced by 3.8 to 4.8\;$\upmu$m simultaneously. 
Interestingly, increasing the upper limit of the power ramp does not positively affect the mode sizes further.
On the contrary best results were obtained by a maximum power level barely surpassing the modification threshold. This is in consistence with our initial deliberations about increasing modification volumes for rising inscription powers (see Fig.~\ref{fig:multipass}).
\par
\begin{figure}[t]
	\includegraphics[width=\linewidth]{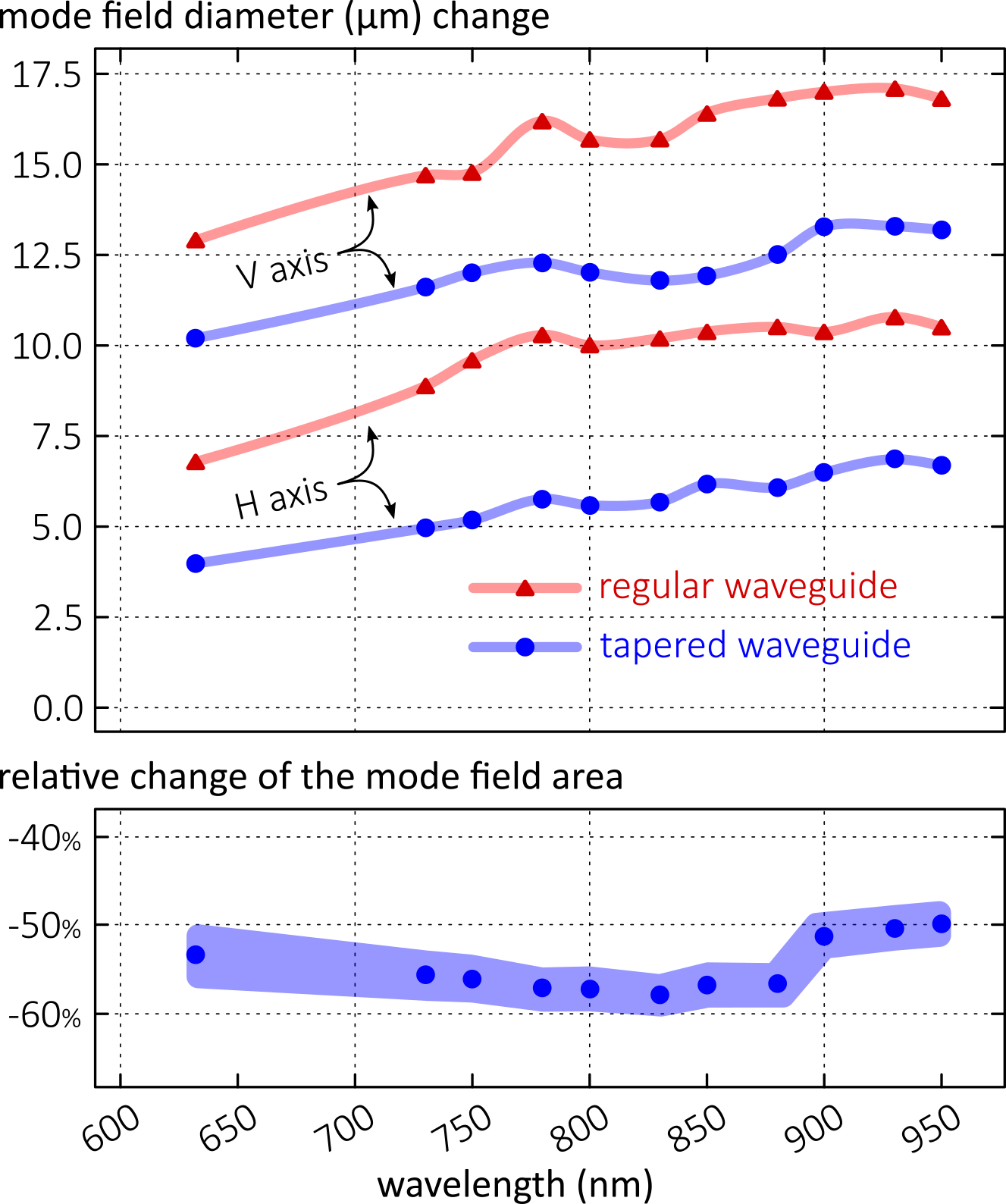}
	\caption{Modification of the mode size for different wavelengths. Top: Itemized components of the mode field diameter for axes H and V regarding a regular waveguide (red triangles) an a tapered waveguide (blue dots). Again, the related line width indicates the error regime of $\pm1.3$\;nm. Bottom: Relative change of the mode field area of a tapered waveguide compared to a regular one. A decrease of as much as 58\% could be achieved. As above the filled out region is linked to the error range. All data was recorded at a power ramp limit $P_\mathrm{a} / P_0 = 66.7\%$.}
	\label{fig:size}
\end{figure}
Following this, the MFD was investigated for a broad range of wavelengths from 632 to 950\;nm (see Fig.~\ref{fig:size}). Therefore, the waveguides were fed by light from a HeNe laser or a tunable Ti:Sa laser in cw mode. 
Whereas the mode profile along the V axis could be strongly decreased but still above the ideal case, the mode profile along the H axis matches the desired size for standard single mode fiber coupling.
Overall, the resulting mode field area of a tapered waveguide could be decreased down to 42\% as of the size of its regular counterpart.
\par
\begin{figure}[t]
	\includegraphics[width=\linewidth]{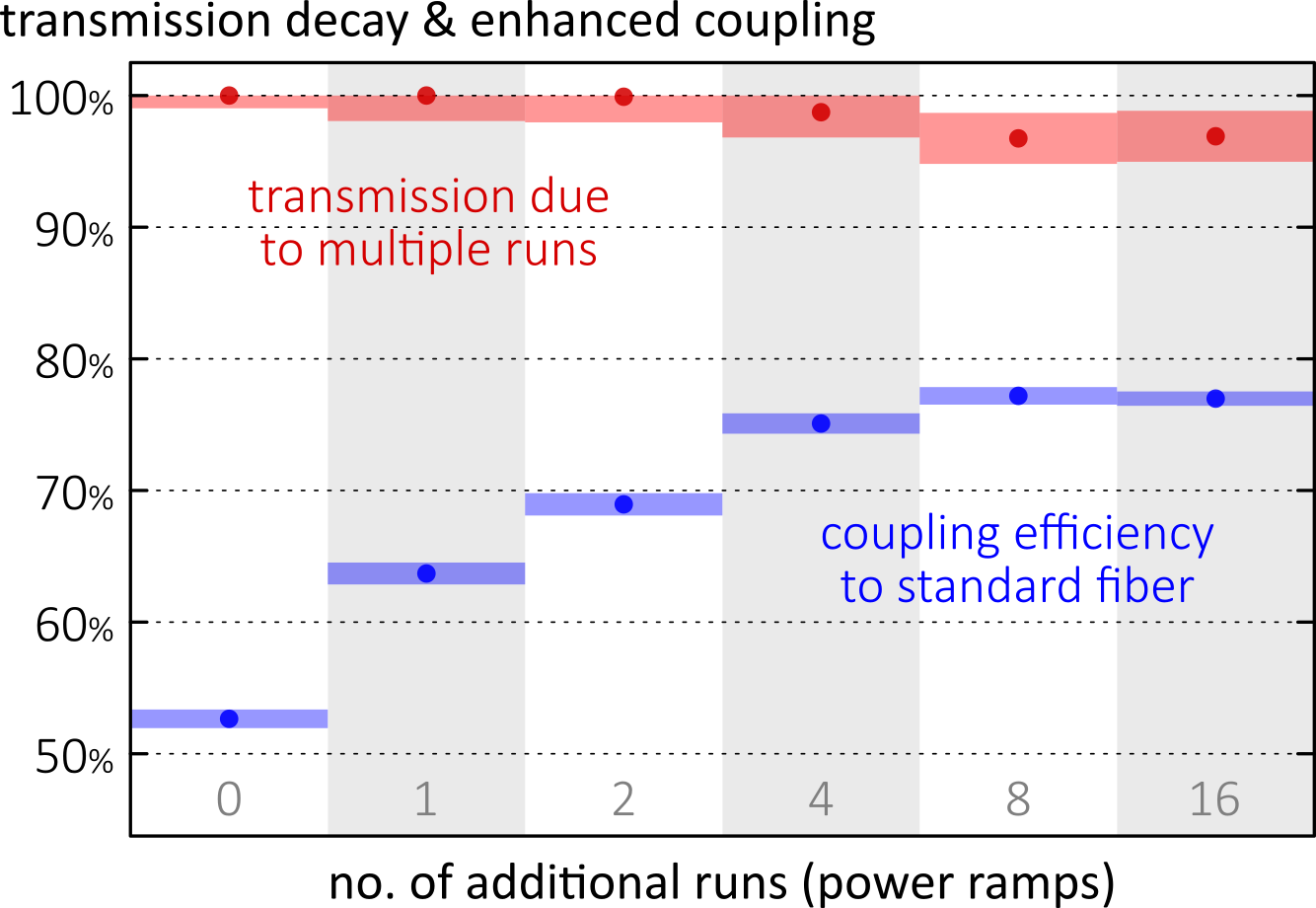}
	\caption{Power transmission (red) and coupling efficiency to a standard fiber (blue) at 800\;nm for different numbers of additional writing sequences. While the coupling efficiency is enhanced from 52\% up to 77\% the transmission compared to a regular waveguide (first column) only drops by 3.5\%. The corresponding filled regions indicate the measurement errors. Again $P_\mathrm{a}$ was set to $66.7\%$ of $P_0$.}
	\label{fig:coupling}
\end{figure}
As discussed before, besides power regulation during tapering, another adjustment can be made by changing the number of repetitions the fs laser writes the taper. Therefore, we iterated the writing process at the waveguide output from 1 to 16 times. In Fig.~\ref{fig:coupling} the impact of additional writing steps on the coupling efficiency $\eta$ described by the overlap integral
\begin{align*}
  \eta = \frac{ \left| \displaystyle\int E_{\mathrm{fiber}}^{\ast} E_{\mathrm{waveguide}} \mathrm{d} A \right| ^{2} }{ \displaystyle\int \left| E_{\mathrm{fiber}} \right| ^{2} \mathrm{d} A \displaystyle\int \left| E_{\mathrm{waveguide}} \right| ^{2} \mathrm{d} A}
\end{align*}
with $A$ as integration area around the mode profile is shown. Based on the MFD of 5.5\;$\upmu$m for standard single mode fibers at the proximity of 800~\;nm, the coupling efficiency could be increased from 52\% for a regular waveguide to 77\% for the best tapered waveguide. A clear saturation effect is visible at a number of repetitions of greater than 4. In parallel, the transmission rate was measured by recording the input and output power of the same 800\;nm laser light as used for the near field imaging. With respect to a regular waveguide, the drop in the transmitted power is negligible (3.5\%), thus confirming the nearly ideal adiabatic modification of the tapered waveguides.
\par
The best result in terms of the overall performance was obtained with a taper implemented by a power ramp rising to a maximum power level just above the threshold for permanent glass modification and written eight times along the end of a regular waveguide. Beside the small writing effort the additional consumption of space can be scaled down to 1\;mm per site.
A further modification towards perfect matching of MFD from fiber to chip may be reached by combining the tapering technique with the shaping of the writing laser focus~\cite{Szameit2007,Salter2012}. In this way the MFD along the V-axis could be shaped down as the size along the H-axis enabling a circular profile and still enable three dimensional structuring.
\par
In conclusion, we demonstrated an efficient method for mode compression of fs laser-written waveguides. Our scheme is attractive since it minimizes the experimental effort and maintains the capability of fabricating precise three dimensional waveguide trajectories. 
While the power level of the additional runs within the tapered region merely shows an impact on the resulting mode profile the number of reruns is substantial for the change of the mode field diameter. 
The demonstrated enhancement of coupling efficiency to standard optical fibers will give a boost on the performance of fs laser-written photonic devices, in particular in the field of on-chip quantum applications.
\section*{Acknowledgements}
The authors gratefully acknowledge financial support from the Deutsche Forschungsgemeinschaft (grants NO462/6-1, SZ 276/7-1, SZ 276/9-1, SZ 276/12-1, BL 574/13-1, GRK 2101/1), the European Commission  EQuaM (No. 323714), PICQUE (No. 608062), QUCHIP (No. 641039) and the Austrian Science Fund (FWF) via PhoQuSi (No. Y585-N20) and CoQuS (No. W1210-N25).

\clearpage

\end{document}